\documentclass[latex,iicol,sn-mathphys-num]{sn-jnl}

\usepackage{graphicx}%
\usepackage{multirow}%
\usepackage{amsmath,amssymb,amsfonts}%
\usepackage{amsthm}%
\usepackage{mathrsfs}%
\usepackage[title]{appendix}%
\usepackage{xcolor}%
\usepackage{chemformula}
\usepackage{textcomp}%
\usepackage{manyfoot}%
\usepackage{booktabs}%
\usepackage{algorithm}%
\usepackage{algorithmicx}%
\usepackage{algpseudocode}%
\usepackage{listings}%
\usepackage{comment}
\usepackage{tabularx}
\usepackage{booktabs}
\usepackage{makecell}
\usepackage{geometry}

\begin{document}
\title[Article Title]{An efficient forgetting-aware fine-tuning framework for pretrained universal machine-learning interatomic potentials}


\author[1]{\fnm{Jisu} \sur{Kim}}
\equalcont{These authors contributed equally to this work.}
\author[1]{\fnm{Jiho} \sur{Lee}}
\equalcont{These authors contributed equally to this work.}

\author[1]{\fnm{Sangmin} \sur{Oh}}
\author[1]{\fnm{Yutack} \sur{Park}}
\author[1]{\fnm{Seungwoo} \sur{Hwang}}
\author[1,2]{\fnm{Seungwu} \sur{Han}}

\author*[3,4]{\fnm{Sungwoo} \sur{Kang}}\email{sung.w.kang@kist.re.kr}
\author*[5]{\fnm{Youngho} \sur{Kang}}\email{youngho84@inu.ac.kr}

\affil[1]{\orgdiv{Department of Materials Science and Engineering}, \orgname{Seoul National University}, \orgaddress{\city{Seoul}, \postcode{08826}, \country{Republic of Korea}}}

\affil[2]{\orgdiv{AI Center}, \orgname{Korea Institute of Advanced Study}, \orgaddress{\city{Seoul}, \postcode{02455}, \country{Republic of Korea}}}

\affil[3]{\orgdiv{Computational Science Research Center}, \orgname{Korea Institute of Science and Technology (KIST)}, \orgaddress{\city{Seoul}, \postcode{02792}, \country{Republic of Korea}}}

\affil[4]{\orgdiv{Division of Nanoscience and Technology, KIST School}, \orgname{University of Science and Technology (UST)}, \orgaddress{\city{Seoul}, \postcode{02792}, \country{Republic of Korea}}}

\affil[5]{\orgdiv{Department of Materials Science and Engineering}, \orgname{Incheon National University}, \orgaddress{\city{Incheon}, \postcode{22012}, \country{Republic of Korea}}}

\abstract{Pretrained universal machine-learning interatomic potentials (MLIPs) have revolutionized computational materials science by enabling rapid atomistic simulations as efficient alternatives to \textit{ab initio} methods. Fine-tuning pretrained MLIPs offers a practical approach to improving accuracy for materials and properties where predictive performance is insufficient. However, this approach often induces catastrophic forgetting, undermining the generalizability that is a key advantage of pretrained MLIPs. Herein, we propose reEWC, an advanced fine-tuning strategy that integrates Experience Replay and Elastic Weight Consolidation (EWC) to effectively balance forgetting prevention with fine-tuning efficiency. Using Li$_6$PS$_5$Cl (LPSC), a sulfide-based Li solid-state electrolyte, as a fine-tuning target, we show that reEWC significantly improves the accuracy of a pretrained MLIP, resolving well-known issues of potential energy surface softening and overestimated Li diffusivities. Moreover, reEWC preserves the generalizability of the pretrained MLIP and enables knowledge transfer to chemically distinct systems, including other sulfide, oxide, nitride, and halide electrolytes. Compared to Experience Replay and EWC used individually, reEWC delivers clear synergistic benefits, mitigating their respective limitations while maintaining computational efficiency. These results establish reEWC as a robust and effective solution for continual learning in MLIPs, enabling universal models that can advance materials research through large-scale, high-throughput simulations across diverse chemistries.}




\maketitle

\section{Introduction}\label{sec1}

Machine-learning interatomic potentials (MLIPs) have significantly advanced computational materials science by accelerating atomistic simulations as efficient surrogates for \textit{ab initio} calculations \cite{MLPreview,BPNN,GAP,NequIP,MACE}. By learning complex potential energy surfaces (PES) from density functional theory (DFT) datasets, MLIPs enable large-scale simulations that retain near-DFT accuracy while reducing computational cost by several orders of magnitude \cite{ML_accel_1,ML_accel_2,ML_accel_CSP}. Initially, MLIPs were developed as bespoke models trained specifically for particular material systems using neural networks or Gaussian approximations with hand-crafted input features, demonstrating excellent predictive performance within narrowly defined chemical spaces \cite{JH_argyrodite, Hong_etching, Choi_thermal, Kang_qd, Jung_Pt, BPNN, GAP, MLPreview}. More recently, substantial efforts have been devoted to developing pretrained MLIPs by leveraging graph neural networks—a deep learning architecture well-suited for atomistic modeling—and training them on large-scale datasets covering most elements in the periodic table, such as the Materials Project \cite{MaterialsProjec}, Alexandria Database \cite{alexandria}, and OMat24 \cite{omat}. Interestingly, although these pretrained MLIPs \cite{sevennet,CHGNet,MACE_MP_0,M3GNet,eSen,Grace} are primarily trained on crystal-structure databases, they exhibit remarkable transferability compared to bespoke models, occasionally extending their applicability even to out-of-domain systems \cite{Kang_GNN}. Examples include inorganic crystals with previously untrained compositions \cite{Roadmap, Matbench}, organic electrolytes \cite{LiOrganic}, and metal-organic frameworks \cite{MACE_MP_0}.

Despite their generalization capabilities, pretrained MLIPs still struggle when applied to materials and configurations that deviate from their training dataset. One of the well-known issues is the systematic softening of the PES; MLIPs tend to underestimate DFT energies and forces of high-energy configurations. This occurs because such pretrained models are predominantly trained on near-equilibrium structures \cite{chgnet_force,surface_energy}. As a result, simulations using pretrained MLIPs exhibit inaccuracies in modeling physical properties associated with high-energy configurations, such as defect formation, ion migration, and vibrational behavior \cite{chgnet_force}. To address this issue, fine-tuning, where a pretrained MLIP is refined using a targeted dataset, has emerged as an effective strategy that can enhance the accuracy of the MLIP model for a specific system with minimal additional effort. \cite{chgnet_force,surface_energy,FT_prl,FT_LayerFreeze,FT_Mace,FT_eqV2}.

\begin{figure*}[h!]
\centering
\includegraphics[width=\textwidth]{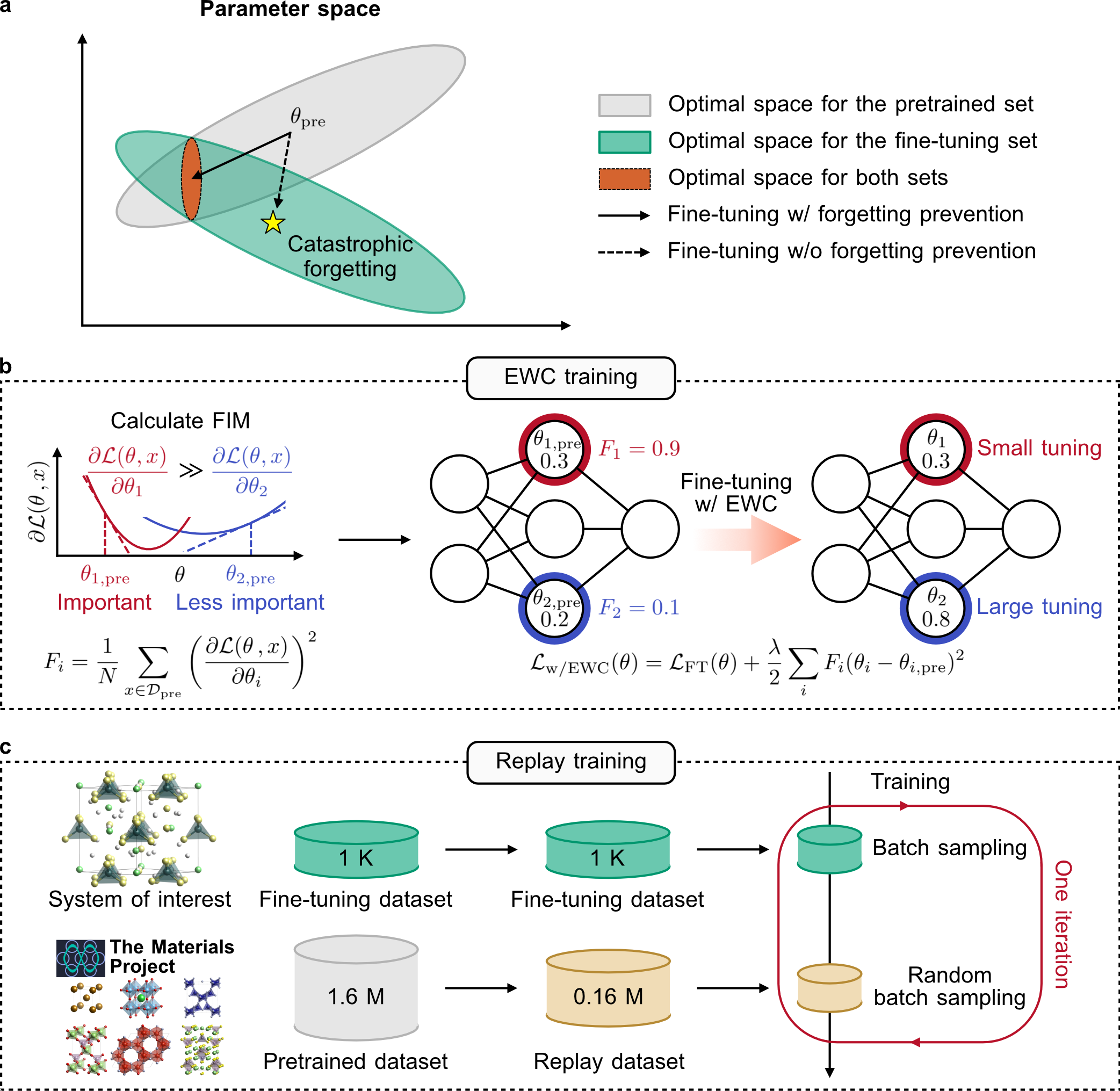}
\caption{\textbf{Schematic overview of forgetting prevention strategies introduced in this work.} \textbf{a} Illustration of catastrophic forgetting in parameter space. Without forgetting prevention, fine-tuning on a new dataset shifts the model parameters away from the optimal region for the pretrained dataset, leading to the loss of previously learned information. \textbf{b} EWC: Importance of pretrained SevenNet-0 parameters is estimated using the Fisher information matrix (FIM), where parameters with high curvature (i.e., high sensitivity in loss) are more strongly regularized during fine-tuning. EWC modifies the fine-tuning loss by penalizing deviation from important pretrained parameters, guiding the model toward a region in parameter space that maintains high accuracy for both the pretrained dataset and the fine-tuning set. \textbf{c} Replay: A Replay set sampled from the pretrained dataset, is used alongside the fine-tuning set during training. In each iteration, a batch is drawn from both the fine-tuning and Replay sets, enabling the model to periodically revisit prior knowledge and reduce forgetting.}
\label{fig:methodologies}
\end{figure*}

However, fine-tuning pretrained MLIPs introduces a fundamental challenge known as catastrophic forgetting, which refers to the loss of knowledge from the pretrained dataset as a result of adjusting model parameters to fit a new dataset (see Fig.~\ref{fig:methodologies}a). This catastrophic forgetting compromises the model’s generalizability—one of the key advantages of pretrained models over bespoke models. For instance, a recent study reported that fine-tuning a pretrained MLIP on dimethyl carbonate liquid configurations enhanced its performance for similar linear-carbonate solvents but degraded its accuracy for chemically distinct cyclic-carbonate systems~\cite{LiOrganic}. Note that preserving the generalizability of MLIPs is critical in materials research, which often demands consistent performance across diverse materials and configurations. In particular, it is essential for large-scale materials discovery, which requires efficient and accurate high-throughput computations across a broad range of material properties over thousands of candidate systems. As such, practical fine-tuning approaches should effectively prevent catastrophic forgetting. 

At the same time, high efficiency and feasibility are also essential requirements for ensuring the broader applicability of fine-tuning methods. When refining an MLIP to improve its prediction of a specific property across a broad range of materials (e.g., Li-ion diffusivity in Li-ion conductors), collecting data for all relevant materials and fine-tuning the model accordingly can be computationally prohibitive due to the high cost of DFT calculations needed to generate extensive training datasets. On the other hand, retraining a model on both the target-specific and original pretrained datasets to mitigate catastrophic forgetting is exhausting and resource-intensive, rendering the approach impractical in many cases. As a result, it is crucial to develop an effective fine-tuning method that enhances accuracy for target systems without sacrificing generalizability, using minimal training data. Such a methodology enables continual learning of pretrained MLIPs, thereby broadening their applicability. 

Several fine-tuning strategies have been proposed in the machine learning field to mitigate catastrophic forgetting in pretrained models, which can be classified into three categories: regularization~\cite{Regularization_EWC,Regularization_SI}, Experience Replay~\cite{Replay_experience1,Replay_experience2}, and architectural modification~\cite{Architecture_PNN,Architecture_dynamic}. Regularization techniques, such as Elastic Weight Consolidation (EWC) \cite{Regularization_EWC}, constrain important model parameters to remain close to their pretrained values. Experience Replay \cite{Replay_experience1,Replay_experience2} (hereafter referred to as Replay) aims to preserve prior knowledge by incorporating a subset of pretrained data into the training set during fine-tuning. Architectural modifications \cite{Architecture_PNN,Architecture_dynamic} mitigate interference between tasks by structurally isolating pretrained parameters using task-specific subnetworks or additional network modules. While these methodologies have proved promise in various machine-learning domains, their applicability to MLIPs remains largely unexplored. Given that molecular dynamics (MD) simulations typically require numerous energy and force evaluations, architectural modification, which involves an increase in model size and complexity, is less suitable for MLIPs because it can incur excessive computational cost for MD simulations.   

Herein, we propose an advanced fine-tuning strategy for pretrained MLIPs that integrates Replay and EWC, referred to as reEWC. To demonstrate its effectiveness, we apply reEWC—alongside Replay and EWC—to fine-tune the pretrained SevenNet-0~\cite{sevennet} model using a fine-tuning set for Li$_6$PS$_5$Cl (LPSC), a representative argyrodite-type Li solid-state electrolyte (SSE) known to suffer from severe PES softening issues~\cite{Pretrained_sulfide}. Our results show that Replay, EWC, and reEWC improve accuracy on LPSC, resolving PES softening and enabling the estimation of Li diffusivity comparable to ground-truth DFT results. In addition, they effectively mitigate catastrophic forgetting and help preserve the generalizability of the pretrained model. This effectiveness is evident when compared to Vanilla, a naive fine-tuning approach without memory retention, which suffers severely from catastrophic forgetting. Interestingly, the MLIP models fine-tuned using the forgetting-aware methods often exhibit improved predictive performance not only for other argyrodite materials with compositions distinct from LPSC, but also for non-argyrodite Li SSEs—including sulfides, oxides, nitrides, and halides~\cite{SSE_review1,SSE_review2,SSE_review3}—despite being fine-tuned solely on the LPSC dataset. This implies that the applied methods enable effective knowledge transfer across chemically diverse systems. However, Replay and EWC each have their limitations. Replay poses a risk of degrading MD performance for materials with chemistries that differ substantially from the target system. EWC, on the other hand, is less effective than Replay in reducing energy and force errors. In contrast, reEWC achieves a synergistic effect, effectively balancing stability—preserving general knowledge across a broad chemical space—and plasticity—enabling accurate learning of new, material-specific behaviors~\cite{Plasticity}. Furthermore, reEWC facilitates knowledge transfer to chemically distinct domains while minimizing the risk of spurious MD behavior, which may occur in MLIP models fine-tuned using Replay alone. 

\section{Results}\label{sec2}

\subsection{Technical backgrounds}\label{subsec1}

In this work, we mainly compare four fine-tuning methods: (1) Vanilla, (2) EWC, (3) Replay, and (4) reEWC. The Vanilla approach is a naive fine-tuning method that relies solely on the fine-tuning dataset, without employing any explicit treatment to prevent catastrophic forgetting—a strategy widely adopted in previous studies to fine-tune pretrained MLIPs~\cite{chgnet_force,surface_energy,FT_prl,FT_LayerFreeze,FT_Mace,FT_eqV2}. Thus, it simply corresponds to continuous learning starting from the pretrained parameters to fit a new dataset. 

EWC is a regularization-based approach that selectively constrains parameter updates to prevent excessive deviation from pretrained parameters (see Fig.~\ref{fig:methodologies}b). Unlike standard $L_2$ regularization, which applies uniform constraints to all parameters, EWC first determines the importance of each model parameter and imposes stronger constraints on those with higher importance. The parameter importance is assessed based on the Fisher information matrix (FIM), which quantifies the sensitivity of model predictions to changes in individual parameters. Herein, due to the complexity of fully considering the FIM (see Methods section), we focus on the diagonal components of the FIM ($F_i$) for each parameter $\theta_i$, which can be computed as follows \cite{Regularization_EWC}:
\begin{equation}
F_i = \frac{1}{N} \sum_{x \in \mathcal{D}_{\mathrm{pre}}} \left( \frac{\partial \mathcal{L}(\theta\,, x)}{\partial \theta_i} \right)^2,
\label{FIM}
\end{equation}
where $\mathcal{L}(\theta\,, x)$ represents the loss function defined over the parameter set $\theta$ and a single data point $x$ from the pretrained dataset $\mathcal{D}_{\mathrm{pre}}$. The partial derivative is evaluated at $\theta = \theta_{\mathrm{pre}}$, where $\theta_{\mathrm{pre}}$ denotes the pretrained parameter set. $N$ corresponds to the number of data points utilized for this computation. The FIM value reflects the curvature of the loss landscape with respect to each model parameter, indicating that minimizing changes in parameters with large FIM values during fine-tuning helps preserve the knowledge acquired during pretraining.  

The loss function in EWC training is defined as a sum of the conventional loss function for fine-tuning set ($\mathcal{L}_{\mathrm{FT}}(\theta)$) and an additional penalty term based on the FIM:
\begin{equation}
    \mathcal{L}_{\mathrm{w/EWC}}(\theta) = \mathcal{L}_{\mathrm{FT}}(\theta) + \frac{\lambda}{2} \sum_i F_i (\theta_i - \theta_{i,\mathrm{pre}})^2.
\label{ewc_loss}
\end{equation}
where $\lambda$ is a hyperparameter controlling the relative significance of the penalty term. We determine its value to effectively learn the fine-tuning set without compromising the memory retention (see Methods section for details). EWC offers clear advantages in terms of scalability and memory efficiency, as it requires neither architectural modifications nor access to the pretrained dataset. By regularizing model parameters, EWC helps preserve the model's generalizability after fine-tuning. However, the learning efficiency on the fine-tuning dataset may be limited if the parameter flexibility is insufficient to capture the new data. Furthermore, due to simultaneous updates of all parameters during training, the omission of off-diagonal components of the FIM—which are computationally prohibitive to evaluate—may undermine the effectiveness of forgetting prevention.

Replay mitigates catastrophic forgetting by retraining the model on a subset of the pretrained dataset (referred to as the Replay set), generated through random sampling, alongside the fine-tuning dataset during the fine-tuning process (see Fig.~\ref{fig:methodologies}c). To implement the Replay method, we adopt a mini-batch training strategy, where each iteration consists of consecutive training on two mini-batches sampled separately from the fine-tuning and Replay datasets. Specifically, a mini-batch is first sampled from the fine-tuning set to compute the loss and update the parameters for learning the target system. Then, another mini-batch of the same size is sampled from the Replay set to perform an additional parameter update, thereby alleviating catastrophic forgetting of the pretrained knowledge. Here, an epoch is defined as a full pass over the fine-tuning data. Since the Replay set is typically much larger than the fine-tuning set, only a fraction of the Replay set is considered within a single epoch. The main advantage of Replay lies in its capability to effectively prevent undesirable atomic energy shifts in materials beyond the target system during fine-tuning, by explicitly incorporating a part of the pretrained training dataset. Nonetheless, unlike EWC, Replay can lead to significant changes in model parameters during fine-tuning. As a result, the fine-tuned model may exhibit serious inaccuracies in describing certain systems. Additionally, the size of the Replay set, critical to the efficiency of forgetting prevention, can be limited due to memory constraints.
 
reEWC is a hybrid approach that combines the Replay and EWC methods. In the present work, the reEWC method is implemented by adopting the mini-batch training process like in the Replay approach while incorporating the EWC regularization term into the loss function during parameter updates. As will be demonstrated below, this method produces a significant synergistic effect, more effectively balancing accuracy improvement for the target system with the preservation of the pretrained model’s generalizability. Additionally, reEWC enables knowledge transfer to chemically distinct domains with minimal risk of spurious MD behavior, which may arise in MLIP models fine-tuned using Replay alone. 

\subsection{Fine-tuning MLIPs on LPSC dataset}\label{subsec2}

We evaluate the effectiveness of fine-tuning methods on Li SSEs, which are known to exhibit significant PES softening in pretrained MLIPs and are also the focus of extensive materials discovery efforts aimed at discovering novel compounds. As the pretrained MLIP to be fine-tuned, we employ SevenNet-0 (version dated 11 July 2024~\cite{sevennetgithub}), which was trained on the MPtrj dataset ~\cite{CHGNet}. Note that SevenNet model provides excellent scalability for large-scale MD simulations using multiple GPUs, offering a distinct advantage over other graph-based MLIP models \cite{sevennet}.

To construct the fine-tuning dataset, we focus on LPSC, one of the most extensively studied argyrodite-type Li SSEs \cite{Argyrodites}. It exhibits fairly high ionic conductivity, approximately 5 mS/cm~\cite{Argrodites_exp}, at room temperature. This feature facilitates efficient sampling of diverse Li-ion hopping events within a limited simulation time, making LPSC an ideal system for generating the fine-tuning dataset aimed at enhancing the predictability of Li diffusion in Li SSE systems. The fine-tuning dataset is constructed from MD trajectories of an LPSC unit cell (52 atoms) obtained from 100 ps simulations conducted at 600 and 1000 K. At each temperature, 500 structures are uniformly sampled at 200 fs intervals, resulting in a total of 1,000 structures. These structures are then randomly split into training and validation datasets at a 9:1 ratio. On the other hand, to construct the Replay set for both Replay and reEWC, we randomly sample 10\% of the MPtrj dataset. The detailed procedure for training the fine-tuned MLIPs is provided in the Methods section.

\subsection{Forgetting behavior of fine-tuned MLIPs}\label{subsec3}

The effectiveness of the fine-tuning methods studied in this work is evaluated from three perspectives: (i) learning of the target system, assessed by the reduction of loss on the fine-tuning dataset; (ii) prevention of forgetting, assessed by the maintenance of loss on the original pretrained dataset; and (iii) preservation of generalizability, assessed by the extent of parameter shifts relative to the pretrained model. We generate a test set by randomly sampling 10\% of the MPtrj dataset (referred to as sMPtrj), which is used to evaluate forgetting prevention across all fine-tuning methods.

\begin{figure*}[h]
\centering
\includegraphics[width=\textwidth]{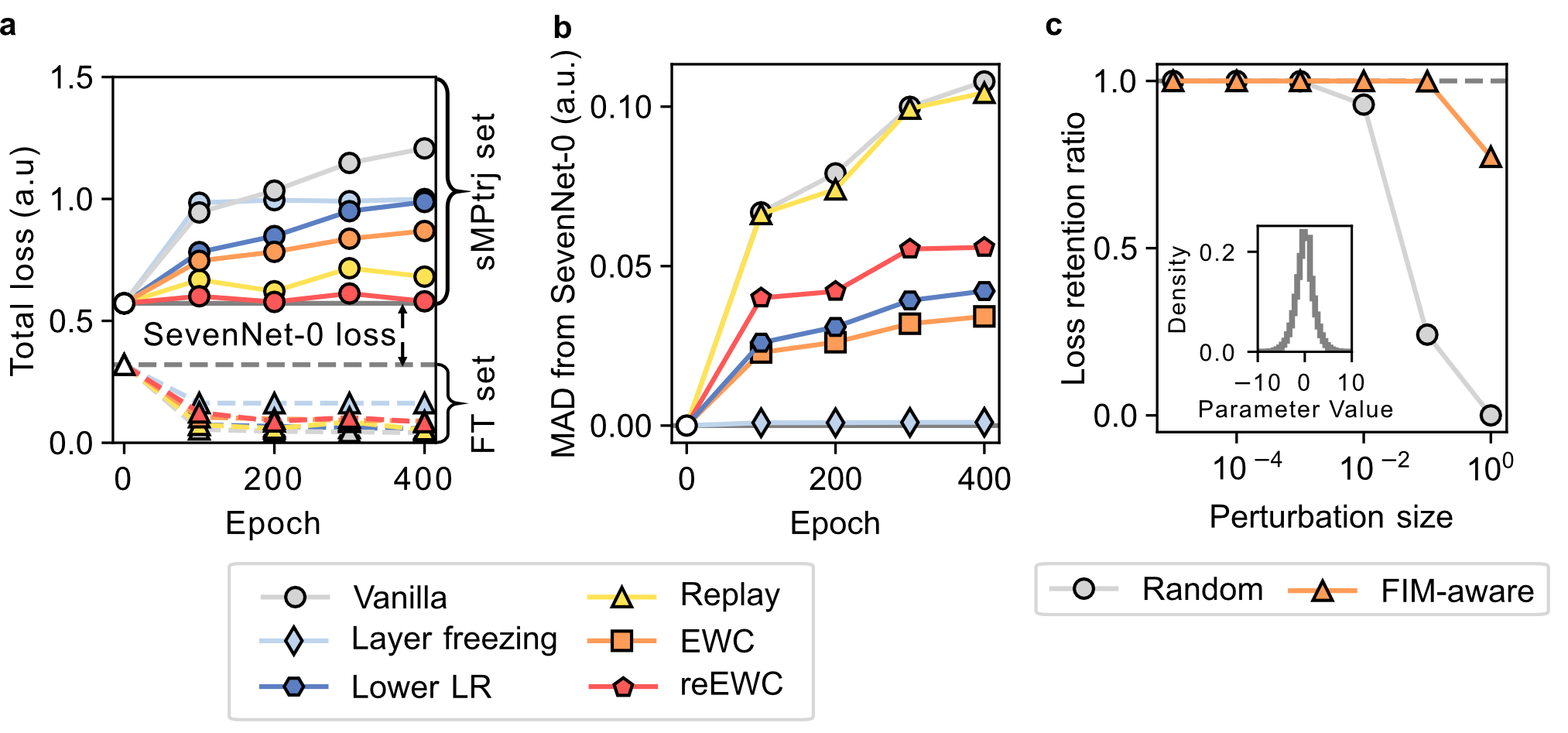}
\caption{\textbf{Training behavior and plasticity-stability analysis of fine-tuned MLIPs.} \textbf{a} Learning and forgetting curves of fine-tuned MLIPs. Total loss over training epochs on the LPSC fine-tuning set (dashed lines, learning) and sMPtrj dataset (solid lines, forgetting) for various fine-tuning methods. Gray lines indicate reference losses from the pretrained SevenNet-0 model. Total loss is defined as \(E_{\mathrm{RMSE}}~(\text{eV/atom}) + F_{\mathrm{RMSE}}~(\text{eV}/\text{\AA}) + 0.01 \times S_{\mathrm{RMSE}}~(\text{kbar})\), where \(E\), \(F\), and \(S\) represent energy, force, and stress, respectively. \textbf{b} Mean absolute distance (MAD) of model parameters from SevenNet-0 over epochs. \textbf{c} Effect of parameter perturbation on model accuracy for SevenNet-0 on the sMPtrj dataset. Two Gaussian noise distribution methods are compared: a random distribution (gray circles) and a FIM-aware distribution (orange triangles). The inset shows the distribution of SevenNet-0 parameter values, concentrated between $-$10 and 10.}
\label{fig:loss}
\end{figure*}

Figure~\ref{fig:loss}a presents the learning curves during fine-tuning over 400 epochs, which are sufficient to ensure convergence of the fine-tuning loss, for both the fine-tuning set (dashed lines) and the sMPtrj dataset (solid lines) across various fine-tuning methods. In this comparison, we also present results from two additional fine-tuning methods, which are well-known variants of the naive Vanilla method: layer-freezing and lower learning-rate (LR) approaches. In the layer-freezing method~\cite{FT_LayerFreeze,LayerFreeze,CHGNet}, only the parameters of the output block are updated, while those of the convolution layers remain fixed. In the lower LR approach~\cite{low_LR1,low_LR2,low_LR3,low_LR4}, the learning rate is set to one-tenth of that used in the other fine-tuning methods. The results indicate that while all methods reduce the loss on the fine-tuning set, the Vanilla, layer-freezing, and lower learning rate approaches suffer from substantial forgetting, resulting in relatively high losses on the sMPtrj dataset. In contrast, forgetting-aware fine-tuning methods such as Replay, EWC, and reEWC yield lower sMPtrj losses, demonstrating their memory retention capabilities. Notably, the reEWC method is highly effective at preventing forgetting, achieving an sMPtrj loss comparable to that of SevenNet-0. 

To gain insight into the origin of forgetting behavior in different fine-tuning methods, we calculate the mean absolute distance (MAD) between the parameters of the fine-tuned models and those of the pretrained SevenNet-0 model, as shown in Fig.~\ref{fig:loss}b. MLIPs fine-tuned using Vanilla and Replay exhibit the most substantial parameter shifts, as neither method imposes explicit constraints on parameter updates. The large parameter shifts observed in the Vanilla method impair memory retention, as it relies solely on the fine-tuning dataset during model refinement. In contrast, Replay incorporates the Replay set during training, guiding parameter updates to mitigate catastrophic forgetting. Nonetheless, the substantial MAD observed in Replay suggests a potential compromise in the generalizability of the fine-tuned model. Indeed, we observe that the Replay-tuned MLIP fails to correctly describe configurations for certain materials outside the training domain, unlike the pretrained model, as will be discussed later. On the other hand, the lower LR method yields parameter changes of similar magnitude to those in the EWC and reEWC methods. However, given the relatively high loss on the sMPtrj dataset, such parameter updates are undesirable with respect to memory retention. By contrast, despite exhibiting similar MADs, the parameter updates in the EWC and reEWC methods more effectively prevent forgetting, thereby achieving a favorable balance between memory retention and learning the target system. As expected, the layer-freezing method results in negligible MADs. However, it yields relatively high losses on both the fine-tuning and sMPtrj datasets, indicating that updating only the output layer parameters is less effective for both learning the new system and preventing forgetting.  

The effectiveness of EWC and reEWC indicates that FIM reliably captures parameter importance—that is, the sensitivity of the pretrained model’s predictions to variations in its parameters. To more evidently demonstrate this point, we conduct a simple numerical experiment by adding noise to the parameter set of the pretrained model and evaluating the resulting performance. Specifically, we generate noise following a Gaussian distribution with a given smearing factor and apply it in two ways. In the first case, referred to as random noise model, the noise is randomly distributed across all model parameters. In the second model, referred to as FIM-aware noise model, the noise for each parameter is scaled proportionally to the inverse Fisher information, $(F + \lambda I)^{-1}$, while maintaining the same total noise magnitude as that of the random noise model. We then evaluate the loss retention ratio on the sMPtrj dataset as a function of the perturbation size, characterized by the Gaussian smearing factor (see Fig.~\ref{fig:loss}c), where the loss retention ratio is defined as:
\begin{equation}
\text{Loss retention ratio} = \frac{\mathcal{L}_{\text{SevenNet-0}}^{\text{sMPtrj}}}{\mathcal{L}_{\text{perturbed}}^{\text{sMPtrj}}}.
\end{equation}
The accuracy of the random noise model on the sMPtrj set rapidly degrades starting from a perturbation size of $10^{-2}$, whereas the FIM-aware noise model maintains its accuracy up to $10^{-1}$. Given that the typical parameter values in SevenNet-0 lie within the range of $[-10, 10]$ (see inset of Fig.~\ref{fig:loss}c), perturbations exceeding $10^{-1}$ are non-negligible. This result suggests that Fisher values effectively encapsulate parameter importance, indicating which parameters should undergo minimal changes during fine-tuning to prevent forgetting.

It is worth noting that reEWC results in noticeably improved forgetting prevention with little loss in accuracy for the target system. Interestingly, reEWC yields lower losses on the sMPtrj dataset than EWC, despite exhibiting larger parameter shifts. This observation suggests that the parameter updates in reEWC are more effective than those in EWC, as the simultaneous application of Replay helps compensate for the limitations of the standard EWC method. Furthermore, reEWC is more robust to the choice of Replay set, compared to Replay. To examine this effect, we train Replay and reEWC models using a Li-focused Replay set, which includes only Li-containing compounds and therefore spans a much narrower chemical space compared to the original Replay set. As shown in Fig. S1, the model fine-tuned using Replay alone exhibits noticeable forgetting on the sMPtrj dataset, likely due to the limited diversity of the Li-focused Replay set. In contrast, the reEWC method effectively preserves the pretrained knowledge, showing minimal increase in sMPtrj loss compared to the original reEWC model trained with the broader Replay set (see Supplementary Note 1 and Fig. S1 for further details). Taken together, these findings clearly demonstrate the synergistic effects of the reEWC method. 

\subsection{PES evaluation of Li SSEs}\label{subsec4}

We evaluate the predictive accuracy of the fine-tuned MLIPs across a range of Li SSEs with diverse compositions and structures. This allows us to assess whether the PES predictions can be improved beyond the fine-tuning domain despite the training dataset being limited to LPSC, and thus, evaluate the extent of knowledge transfer enabled by each fine-tuning method. To this end, we construct test sets from two material groups: (i) argyrodite-type materials, which share a similar configurational space with LPSC, and (ii) non-argyrodite-type materials, which span a broader chemical and structural diversity beyond the fine-tuning domain. For the argyrodite-type test set, we consider compounds with the chemical formula \ch{Li_{24+$x$}$M$4S20$X$4}, where $M$ represents tetravalent (Si, Ge, or Sn) and pentavalent (P, As, or Sb) cations, and $X$ is a halide anion such as Cl, Br, or I. In addition, we account for high-entropy systems in which more than two different non-Li cations and halides are incorporated (e.g., \ch{Li_{26}P2Si2S20Cl2I2}). In this case, assuming that S and $X$ are S$^{2-}$ and $X^-$ ions, respectively, the number of Li atoms is adjusted to maintain charge neutrality. The combination of non-Li cations and halides yields a total of 1,890 possible compositions. Among them, considering the high computational cost of DFT calculations to produce reference data for comparison, we select 126 representative compounds, which cover diverse chemistries, for the test set (see Table S1 for detailed compositions). Note that the argyrodite structure contains two distinct anion sites, 4a and 4c, which can be shared by S and $X$ atoms (see Fig. S2). The site occupancy of S and $X$ on these sites can vary depending on experimental conditions, resulting in structural disorder. In the argyrodite test set, five levels of disorder between S and $X$ (0\%, 25\%, 50\%, 75\%, and 100\% $X$@4c) are randomly applied to the 126 compounds, where $X$@4c indicates the percentage of 4c sites occupied by $X$ atoms. For non-argyrodite types, the test set include nine well-known Li SSE materials: oxides (\ch{Li7La3Zr2O12}; LLZO, \ch{Li_{1.33}Ti_{1.67}Al_{0.33}(PO4)3}; LATP) \cite{llzo,statistical}, sulfides (\ch{Li10GeP2S12}; LGPS, \ch{Li7P3S11}; LPS) \cite{lgps,lps}, halides (\ch{Li3YCl6}; LYC, \ch{Li3YBr6}; LYB) \cite{halide}, nitrides (\ch{Li3N}, \ch{Li9S3N}; LSN) \cite{li3n,li9s3n}, and hydrides (\ch{LiBH4}) \cite{lbh}. 

To generate a specific test dataset, we first optimize the structures of the selected materials at 0 K and then perform \textit{ab initio} molecular dynamics (AIMD) simulations for 100 ps under the NVT ensemble. Each system is gradually heated from 300 K to 1200 K to probe the PES across the temperature range typically used for simulating Li diffusion in SSEs. Test configurations are sampled every 200 fs, resulting in 500 structures per material.

\begin{figure*}[h]
\centering
\includegraphics[width=\textwidth]{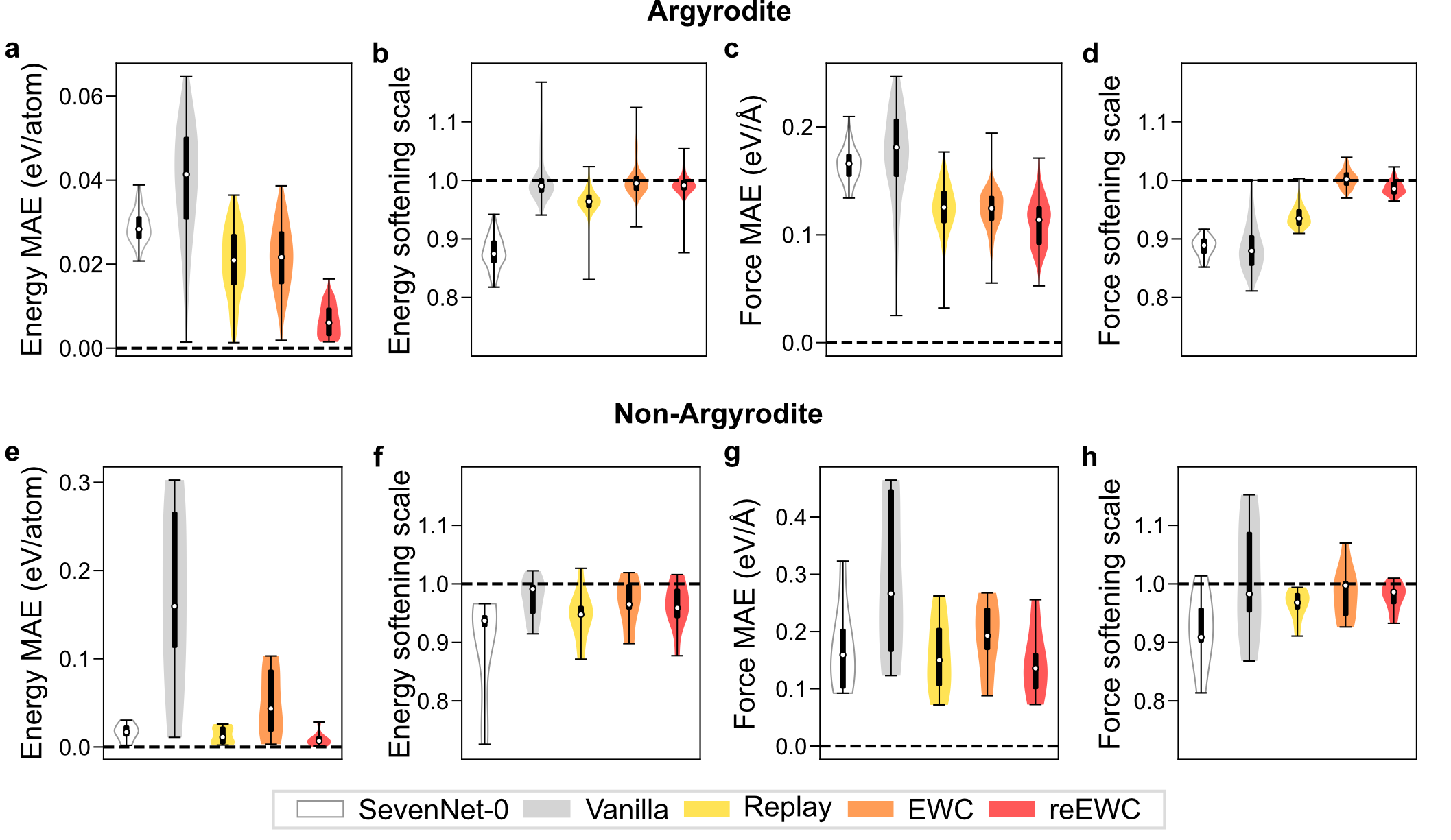}
\caption{\textbf{Energy and force prediction accuracy of MLIPs on Li SSEs AIMD benchmark dataset.} The top row (\textbf{a–d}) presents results for the 126 argyrodite-type test materials, while the bottom row (\textbf{e–h}) shows results for the 9 non-argyrodite-type materials. \textbf{a, e} show energy MAE, \textbf{b, f} energy softening scale, \textbf{c, g} force MAE, and \textbf{d, h} force softening scale. Colors represent different models: white for SevenNet-0, gray for Vanilla MLIP, yellow for Replay MLIP, orange for EWC MLIP, and red for reEWC MLIP. The softening scale, defined as the slope of the linear regression line comparing MLIP predictions to DFT reference values, indicates softening when less than 1 and stiffening when greater than 1. In each violin plot, white dots denote the median, and black bars indicate the interquartile range (25th to 75th percentile).}\label{fig:ef_violin}
\end{figure*}

\begin{figure*}[h]
\centering
\includegraphics[width=\textwidth]{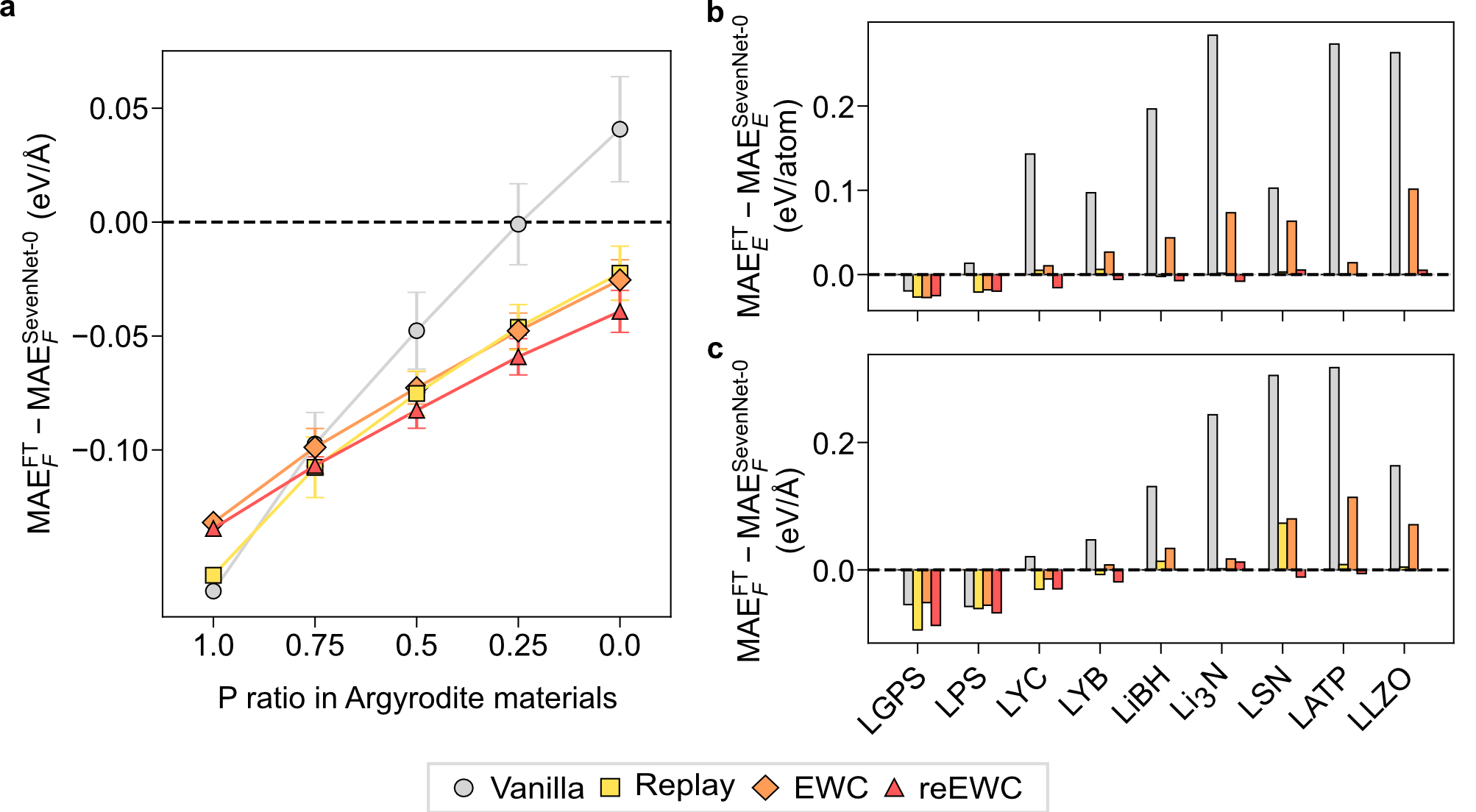}
\caption{\textbf{Assessment of knowledge transfer with respect to the P ratio in argyrodite materials and material type in non-argyrodite systems.} \textbf{a} Differences in the mean absolute force errors between fine-tuned MLIPs ($\rm{MAE}^{\rm{FT}}_{\it F}$) and SevenNet-0 ($\rm{MAE}^{\rm{SevenNet-0}}_{\it F}$) as a function of the P ratio in the argyrodite test set. A P ratio of unity corresponds to LPSC. {\textbf{b,c}} Differences in the mean absolute force and energy errors between fine-tuned MLIPs and SevenNet-0 across nine non-argyrodite-type materials. Error bars represent standard deviations. Gray circles, yellow squares, orange diamonds, and red triangles indicate data for MLIPs fine-tuned using Vanilla, Replay, EWC, and reEWC, respectively.  }\label{fig:p_ratio}
\end{figure*}

Figures~\ref{fig:ef_violin}a–d present the energy and force mean absolute errors (MAEs) on the argyrodite-type test set, along with the corresponding softening scales, for SevenNet-0 and fine-tuned MLIPs. Here, the softening scale is defined as the slope of the MLIP-versus-DFT curves for energy and force, determined via linear fitting (see Fig. S3). A softening scale below 1 indicates softening, whereas a value above 1 indicates stiffening. SevenNet-0 exhibits systematic softening in both energy and force predictions, with softening scales consistently below unity, aligning with previous observations \cite{chgnet_force}. The MLIP trained using the Vanilla method shows significant increases in both energy and force MAEs compared to the reference SevenNet-0 model, indicating a severe forgetting issue. In contrast, the MLIPs fine-tuned using the Replay, EWC, and reEWC hybrid methods do not exhibit such forgetting behavior. Notably, these methods lead to substantial improvements in the predictions of energy, force, and their softening scales, rather than merely preserving the error levels of SevenNet-0. This suggests that the knowledge acquired through fine-tuning on the LPSC dataset effectively transfers to other argyrodite-type systems. We notice that the knowledge transfer becomes more pronounced in systems with higher P ratios, leading to lower force MAEs, as shown in Fig.~\ref{fig:p_ratio}a. This can be attributed to the fact that PS$_4$ units in LPSC, the target material used for fine-tuning, forms a structural backbone influencing the PES associated with Li-ion motion. Consequently, systems containing PS$_4$ units benefit more significantly from fine-tuning. A similar analysis is performed with respect to the halide elements, revealing only a weak correlation (see Fig. S4). Accuracy improvements compared to the reference SevenNet-0 model are observed even at a zero P ratio for the Replay-, EWC-, and reEWC-tuned MLIPs, further demonstrating the knowledge-transfer capability of these forgetting-aware fine-tuning methods. It is noted that among the three approaches, the reEWC hybrid method demonstrates the most balanced and robust performance across all metrics, further highlighting the synergistic effect of combining Replay and EWC strategies.

Figures~\ref{fig:ef_violin}e--h show the test results for the non-argyrodite-type test set. As demonstrated for argyrodite materials, SevenNet-0 exhibits systematic softening in both energy and force predictions, while the Vanilla-tuned MLIP suffers from catastrophic forgetting, resulting in significantly increased energy and force MAEs. In contrast, the Replay, EWC, and reEWC methods largely reduce the energy and force MAEs and improve the softening scales, indicating their effectiveness in enhancing predictive accuracy even for materials outside the fine-tuning domain. In Figs.~\ref{fig:p_ratio}b,c, we quantify the MAE changes of the fine-tuned models relative to SevenNet-0 for each non-argyrodite material. LGPS and LPS exhibit notable decrease in the MAEs, whereas the MAEs for the other materials remain similar or show slight increases compared to those of SevenNet-0. Given both LGPS and LPS contain PS$_4$ units, the knowledge acquired through fine-tuning based on LPSC is effectively transferable to these materials. Among the three forgetting-aware methods, EWC results in relatively larger increases in energy MAEs for the tested nitrides and oxides, which are chemically and structurally more distinct from LPSC than the other non-argyrodite materials. This can be attributed to the lack of consideration of the pretrained dataset in the EWC method, which alters ad-hoc atomic-energy mapping~\cite{mapping} of SevenNet-0 that is already reasonably accurate for such systems. In contrast, this issue is effectively mitigated by the Replay and reEWC methods. It is noted that the force MAEs of the EWC-tuned MLIP still remain low within 0.1 eV/\AA\ of those from SevenNet-0. This suggests that its relatively large energy MAEs mainly arise from a nearly constant shift in atomic energies, allowing the potential gradients to remain largely unchanged. Accordingly, the EWC-tuned model can still reasonably describe the dynamical properties of the nitrides and oxides (see Section~\ref{subsec5}).   Consistent with the results on the argyrodite test set, the reEWC-tuned model demonstrates the best overall performance on the non-argyrodite test set. 

The effectiveness of the forgetting-aware methods is further evaluated through an element-specific accuracy analysis (see Figs. S5a,b). The Vanilla-tuned MLIP shows improvement only for elements present in the fine-tuning set (Li, P, and Cl), while its performance degrades significantly for other elements—particularly metal cations (As, Sb, Si, Ge, and Sn) and even S. While EWC and Replay largely help improve or maintain accuracy, they still lead to degradation for certain elements. In contrast, reEWC enhances accuracy across nearly all elements, confirming its superior effectiveness.

\subsection{Dynamical properties}\label{subsec5}

\begin{figure*}[h]
\centering
\includegraphics[width=\textwidth]{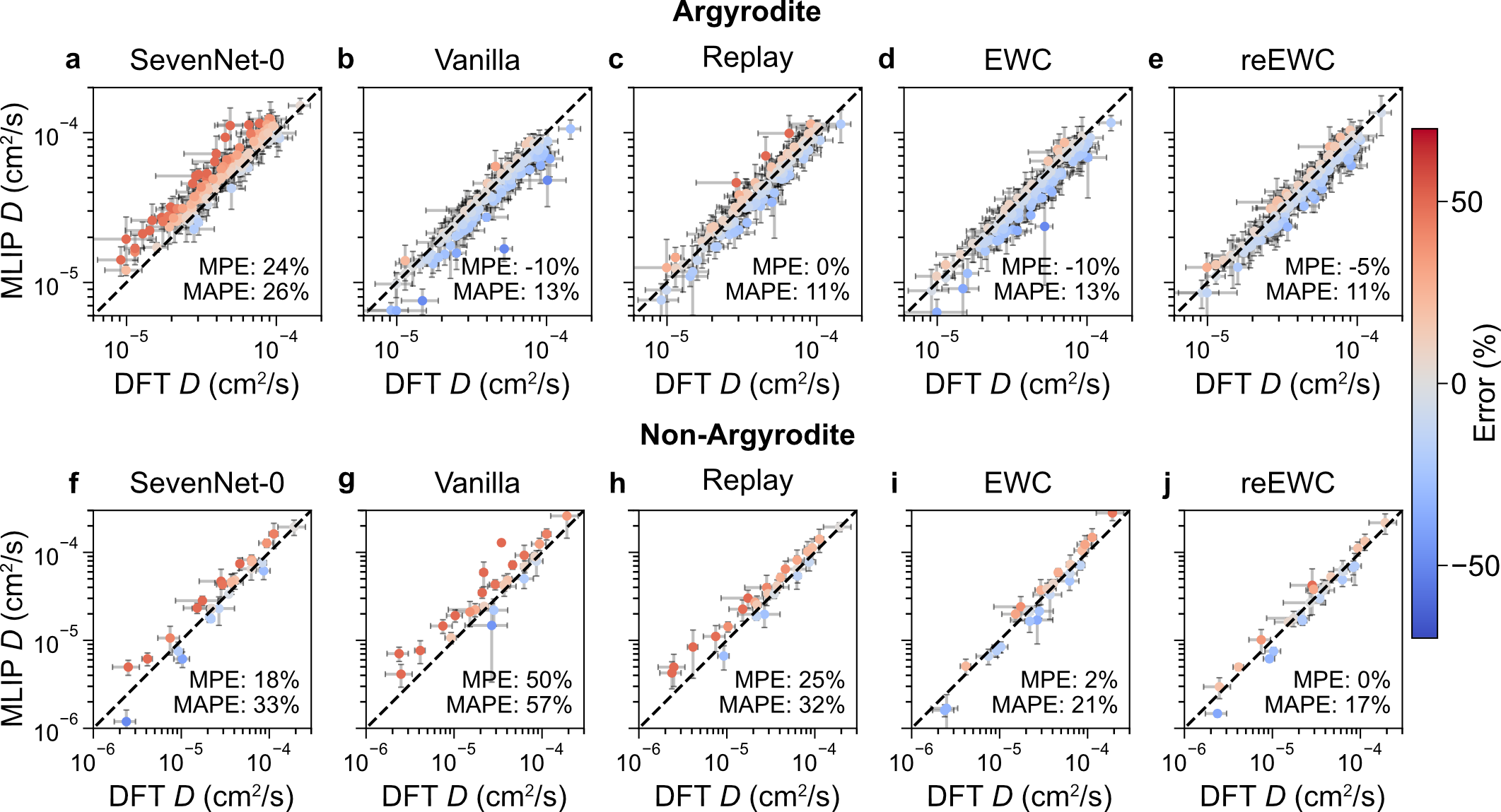}
\caption{\textbf{Evaluation of Li-ion diffusivity prediction of MLIPs.} \textbf{a–e} Parity plots comparing MLIP-predicted Li-ion diffusivities with DFT references across 45 argyrodite-type materials. \textbf{f–j} Corresponding results for 9 non-argyrodite-type materials. (\textbf{a, f}), (\textbf{b, g}), (\textbf{c, h}), (\textbf{d, i}), and (\textbf{e, j}) correspond to SevenNet-0, Vanilla, Replay, EWC, and reEWC MLIPs, respectively. Each point represents the average diffusivity from five independent MD simulations with different initial velocities, with error bars indicating the standard deviation. MPE and MAPE are the mean percentage error and mean absolute percentage error, respectively. The color of each point encodes the percentage error relative to DFT, ranging from $+70\%$ (red) to $-70\%$ (blue).}\label{fig:diff_scatter}
\end{figure*}

So far, we have evaluated the predictive capability of the fine-tuned MLIPs for DFT energies and forces based on snapshots from AIMD trajectories. As a next step, we explicitly perform MD simulations using the fine-tuned MLIPs to assess whether they accurately describe the PES governing atomic motion at finite temperatures, and thereby capture the associated dynamical properties. To this end, we evaluate the diffusivity of Li ions by calculating mean-squared displacements (MSDs) at 800, 1000, and 1200 K, and compare the results with DFT benchmarks, as shown in Fig.~\ref{fig:diff_scatter}. Due to the high computational cost of DFT, we focus on 45 out of the 126 argyrodite-type systems in the test set (see Fig. S6 and Table S1 for detailed compositions), selected to ensure chemical diversity, along with 9 non-argyrodite-type materials. In Fig.~\ref{fig:diff_scatter}, the mean percentage error (MPE) indicates the tendency of MLIPs to overestimate or underestimate diffusivity, thereby reflecting the degree of PES softening. The mean absolute percentage error (MAPE) quantifies the size of the deviation from the reference DFT values.

For argyrodite-type materials, SevenNet-0 systematically overestimates diffusivities, leading to large positive MPE and MAPE values (see Fig.~\ref{fig:diff_scatter}a). This aligns with the PES softening behavior of the pretrained MLIP. All fine-tuning methods examined in this study improve diffusivity predictions, reducing both MPE and MAPE. This observation demonstrates that knowledge learned from the LPSC training set can effectively enhance the PES description of the fine-tuned MLIPs for other argyrodite systems, and consequently, their dynamical properties. Note that although the Vanilla-tuned model appears to reasonably estimate Li migration, this does not indicate effective forgetting prevention and knowledge transfer. Instead, the model is likely overfitting to the LPSC training set, resulting in element-wise imbalances in error (see Fig. S5). Consequently, it exhibits significant inaccuracies in other dynamical properties, which will be discussed later. 

On the other hand, the performance of the fine-tuned models varies significantly for non-argyrodite-type materials. Fine-tuning with the Vanilla method results in substantial increases in both MPE and MAPE compared to SevenNet-0, which already tends to overestimate diffusivities, primarily due to catastrophic forgetting. In contrast, the Replay method maintains errors comparable to those of SevenNet-0. Interestingly, the EWC-tuned model yields lower MPE and MAPE than the Replay-tuned model, highlighting the importance of parameter regularization (see Fig.~\ref{fig:loss}b) in improving the representation of the PES shape (i.e., potential gradients) for materials outside the fine-tuning domain. Most notably, the reEWC hybrid method achieves the lowest errors. 

\begin{figure*}[h]
\centering
\includegraphics[width=\textwidth]{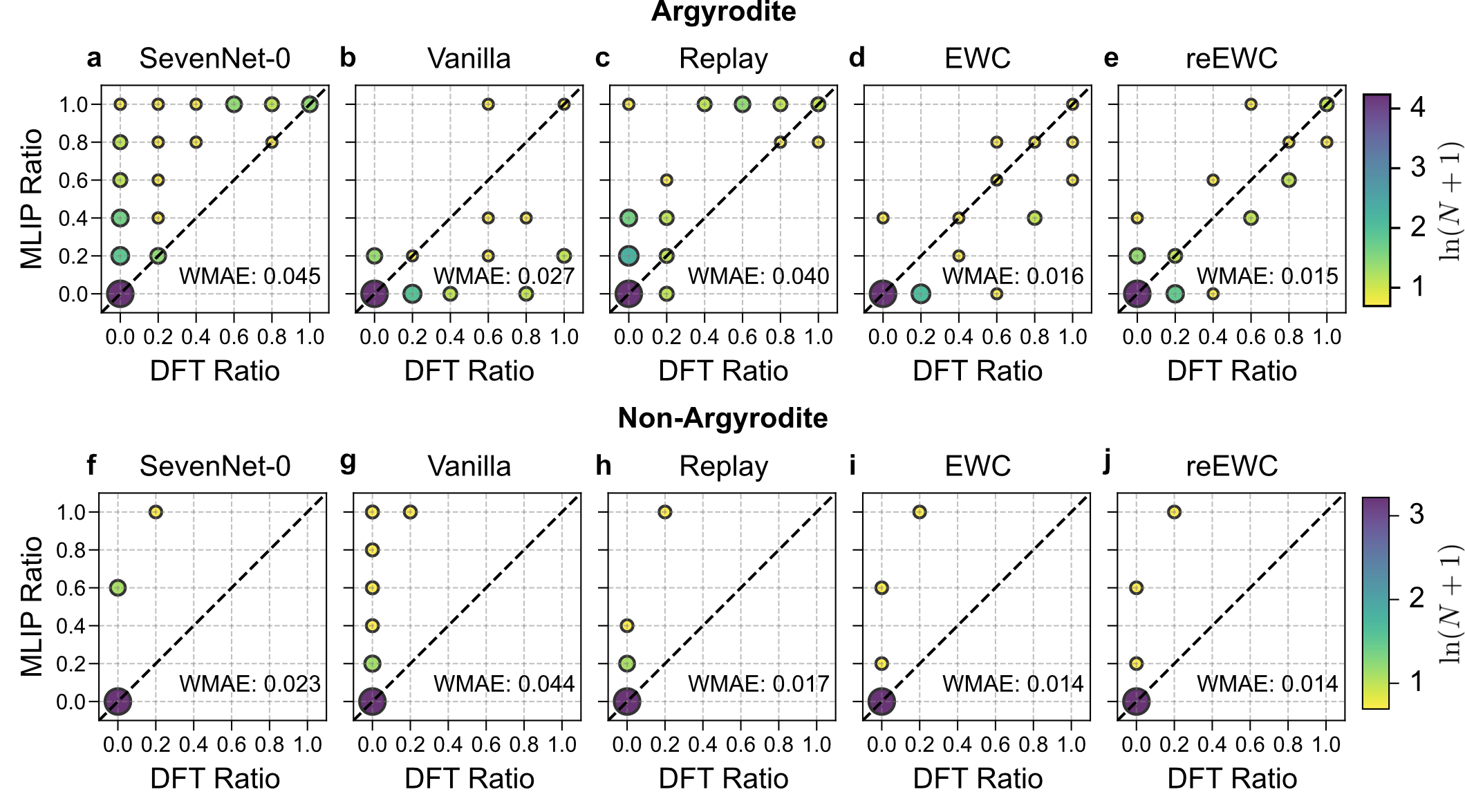}
\caption{\textbf{Consistency of MLIP-MD and AIMD structural stability predictions.} Each panel displays a bubble plot comparing the DFT quasi-melting ratio ($x$-axis) to the MLIP-predicted quasi-melting ratio ($y$-axis). Panels \textbf{a--e} show results for 45 argyrodite-type materials across 90 data points, while panels \textbf{f--j} show results for 9 non-argyrodite-type materials across 18 data points, with each material simulated at two different temperatures (1000 and 1200 K) using five independent MD trajectories each. (\textbf{a, f}), (\textbf{b, g}), (\textbf{c, h}), (\textbf{d, i}), and (\textbf{e, j}) correspond to SevenNet-0, Vanilla, Replay, EWC, and reEWC MLIPs, respectively. The size and color of each point indicates the number of structures exhibiting quasi-melting behavior. The dashed diagonal line (\( y = x \)) denotes perfect agreement between MLIP and DFT predictions. Deviation from this line reflects the extent to which the MLIP model either overpredicts (above the line, softening) or underpredicts (below the line, stiffening) quasi-melting behavior relative to DFT.}\label{fig:melt}
\end{figure*}

In addition to Li diffusion, the ability to maintain the structural framework, associated with the displacement of non-Li ions, at high temperatures is a critical property of Li SSEs that MLIPs should accurately describe. In AIMD simulations for certain materials such as \ch{Li_{6}As_{0.75}Sb_{0.25}S_{5}Br_{0.5}I_{0.5}}, we observe significant displacements of non-Li ions beyond their typical vibrational range at temperatures above 1000 K, resulting in the collapse of their structural frameworks—a phenomenon we refer to as quasi-melting. To evaluate the consistency of quasi-melting behavior between AIMD and MLIP-MD simulations, we compute the quasi-melting ratio for each MD scheme. This ratio is defined as the fraction of quasi-melting occurrences observed in five independent MD simulations conducted at 1000 and 1200 K. Quasi-melting is considered to occur when all non-Li ions exhibit a MSD rate exceeding 0.36 \AA$^2$/ps. This criterion implies that such elements reach second- or third-nearest-neighbor sites within 100 ps. Figures~\ref{fig:melt}a--e compare the quasi-melting ratios between MLIP and DFT simulations for the 45 argyrodite-type materials, with weighted mean absolute errors (WMAEs) (see Methods section for details on WMAE calculations). SevenNet-0 tends to overpredict quasi-melting even for materials that remain structurally stable in AIMD simulations due to PES softening. In contrast, the Vanilla-tuned MLIP tends to suppress quasi-melting relative to DFT, indicating the presence of PES stiffening. This stiffening is likely a result of the model being overly tuned to LPSC, which remains structurally stable in AIMD simulations. The Replay method offers slight improvement; however, the fine-tuned MLIP still tends to overpredict quasi-melting, similar to SevenNet-0. On the other hand, the EWC- and reEWC-tuned MLIPs yield quasi-melting ratios comparable to DFT results, achieving the lowest WMAEs and demonstrating superior accuracy. 

\begin{figure}[h!]
\centering
\includegraphics[]{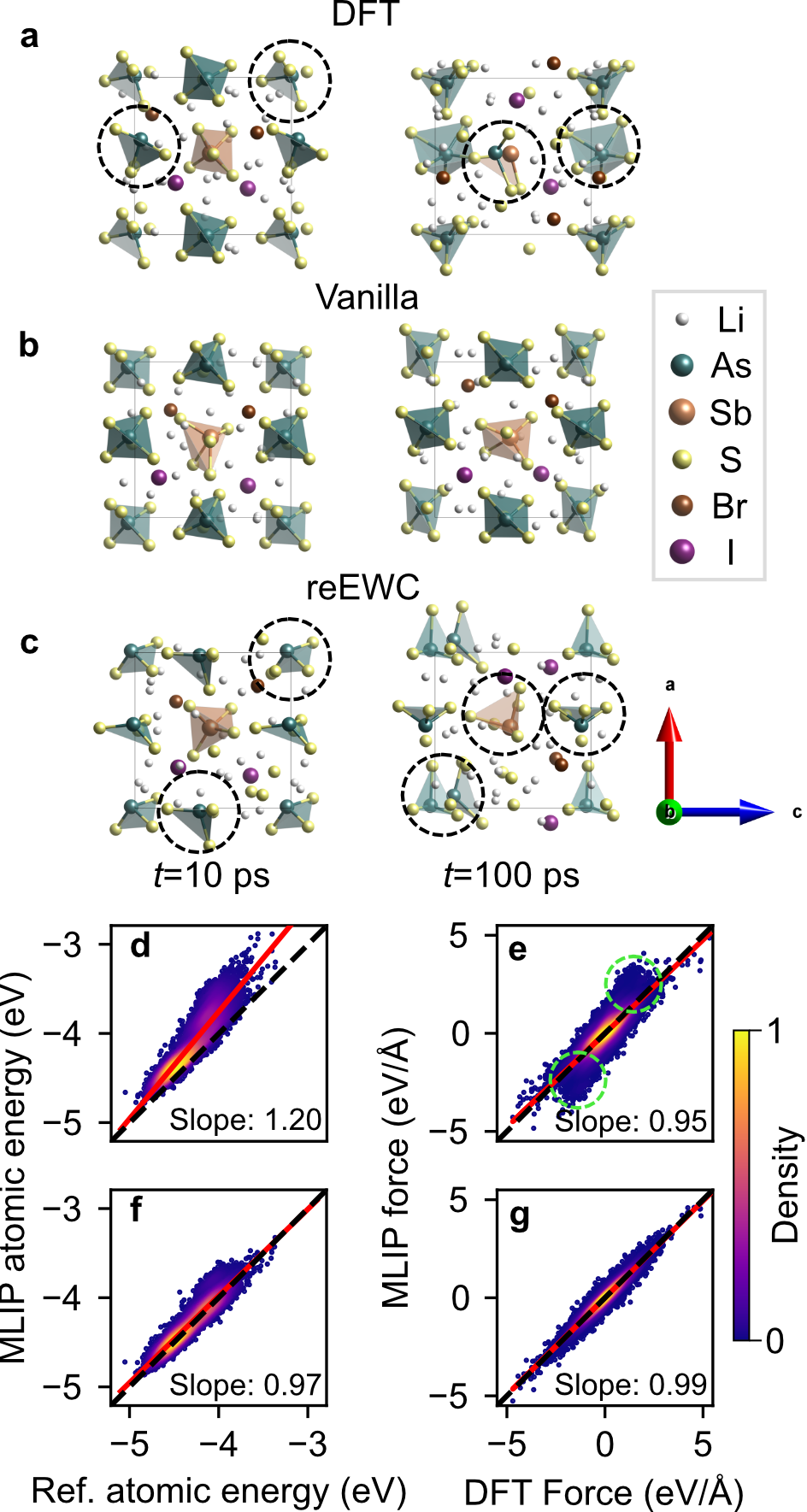}
\caption{\textbf{Dynamical behavior of \ch{Li_{6}As_{0.75}Sb_{0.25}S_{5}Br_{0.5}I_{0.5}}.}
\textbf{a} DFT-MD, \textbf{b} Vanilla MLIP, and \textbf{c} reEWC MLIP, respectively. Distorted metal–S$_4$ tetrahedra are highlighted with dashed circles, reflecting structural disorder. \textbf{d–g} Parity plots of sulfur (S) atomic energy and atomic force predictions for \textbf{d,e} Vanilla MLIP, and \textbf{f, g} reEWC MLIP. Predictions are sampled from a 100 ps AIMD trajectory at 1200 K, with reference atomic energies calculated using a reference MLIP trained on MD trajectories of \ch{Li_{6}As_{0.75}Sb_{0.25}S_{5}Br_{0.5}I_{0.5}} at 800, 1000, 1200, and 1500 K. Red lines in each plot indicate the slope, with values displayed in the lower-right corner.}\label{fig:mapping}
\end{figure}

As an illustrative example of PES stiffening induced by the Vanilla method, we present MD snapshots of \ch{Li_{6}As_{0.75}Sb_{0.25}S_{5}Br_{0.5}Cl_{0.5}} at 1200 K (see Fig.~\ref{fig:mapping}). In AIMD simulations, the structure undergoes clear deformation after 100 ps due to significant non-Li ion displacements, particularly around the AsS$_4$ and SbS$_4$ tetrahedra (highlighted by dotted circles in Fig.~\ref{fig:mapping}a). In contrast, the initial structural motifs remain largely intact when simulated with the Vanilla-tuned MLIP (see Fig.~\ref{fig:mapping}b). For this material, fine-tuned MLIPs employing the forgetting-aware methods successfully reproduce the quasi-melting behavior observed in the AIMD calculations, as shown in Fig.~\ref{fig:mapping}c for the reEWC-tuned model. 

As discussed above, PES stiffening of the Vanilla-tuned MLIP is likely due to overfitting to LPSC. To support this, we evaluate its atomic energies and forces on sulfur, a constituent of LPSC, using AIMD trajectories that include collapsed $M$S$_4$ units, as shown in Figs.~\ref{fig:mapping}d,e. Because atomic energies cannot be directly obtained from DFT calculations, the reference atomic energies in Fig.~\ref{fig:mapping}d are taken from the MLIP fine-tuned exclusively on MD trajectories of \ch{Li_{6}As_{0.75}Sb_{0.25}S_{5}Br_{0.5}Cl_{0.5}} over the temperature range of 800–1500 K. The Vanilla-tuned MLIP clearly overestimates the reference atomic energies of high-energy structures with collapsed $M$S$_4$ units, indicating PES stiffening. As a result, this model exhibits an excessive preference for preserving the original structural framework, compared to DFT. Although its average force softening scale, defined as the slope, is 0.95 (see Fig.~\ref{fig:mapping}e), a large portion of forces exceed the magnitudes of the corresponding DFT forces (highlighted by the green dashed circle in Fig.~\ref{fig:mapping}e), consistent with the overestimated atomic energies. We also check the energy and force parity plots for Sb and As, which are not present in LPSC. As expected, these elements rarely experience stiffening issues (see Fig. S7). In contrast, the reEWC-tuned MLIP exhibits excellent agreement with the reference values for both atomic energy and force, as shown in Figs.~\ref{fig:mapping}f,g, respectively.

\begin{figure*}[h!]
\centering
\includegraphics[width=\textwidth]{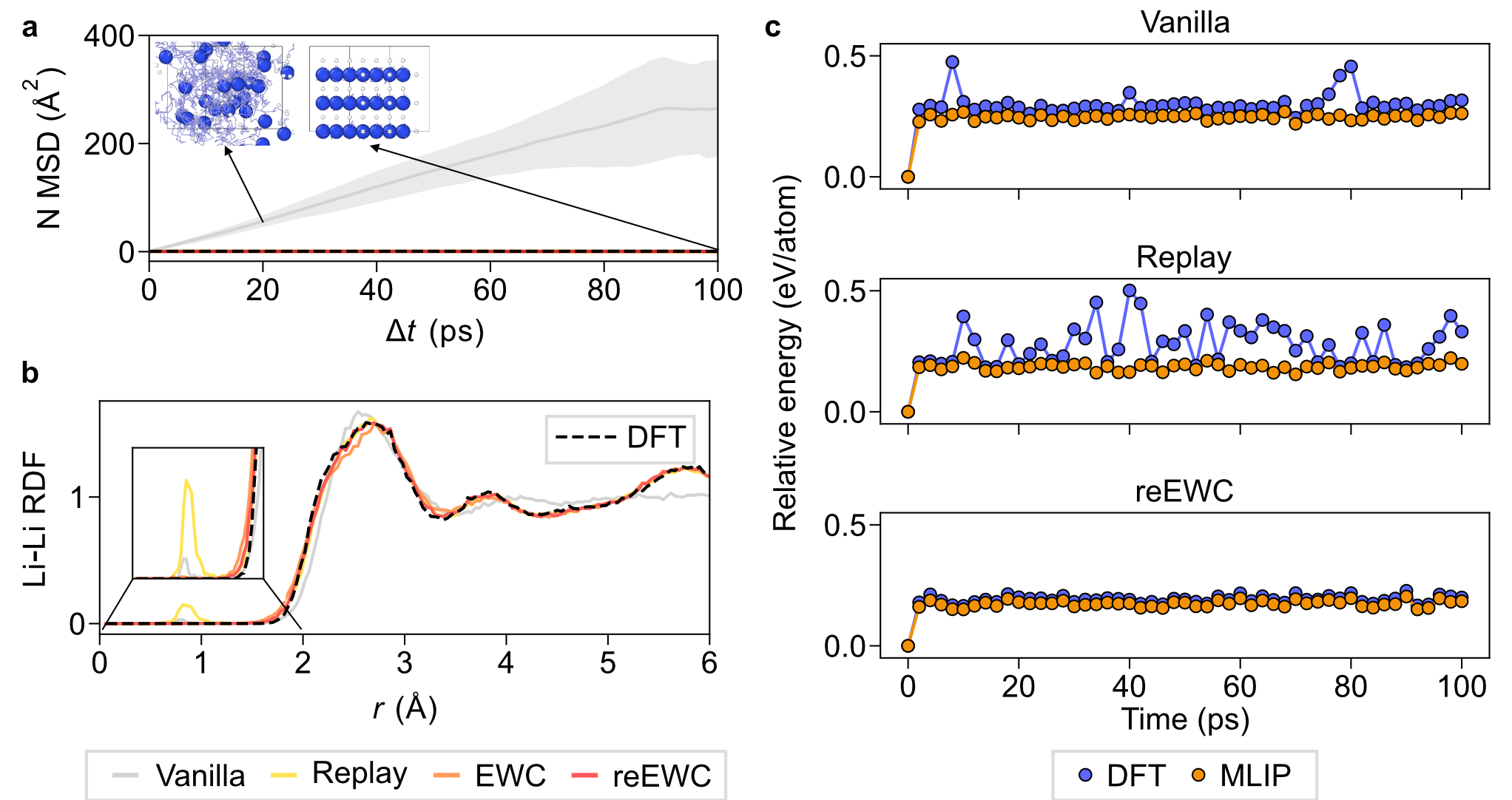}
\caption{\textbf{Dynamic behavior of \ch{Li3N}.}
\textbf{a} Time-averaged mean-squared displacement (MSD) for N atoms at 1200 K. Solid lines represent averages over five simulations, with shaded regions indicating standard deviations. Insets show snapshots from Vanilla-tuned MLIP MD (left, 20~ps) and DFT-MD (right, 100~ps), with N atom trajectories highlighted in blue. \textbf{b} Li--Li radial distribution function (RDF), averaged over snapshots from the 1200 K MD simulations. The inset zooms in on the 0--2~\AA\ region, highlighting a peak near 1~\AA\ that corresponds to unphysical Li--Li bonding in simulations using the Vanilla- and Replay-tuned MLIPs. Black dotted lines indicate DFT reference data. \textbf{c} Relative energies of structures sampled from MD trajectories using the Vanilla-, Replay-, and reEWC-tuned MLIPs, evaluated through DFT single-point calculations. Fifty structures were extracted at 2~ps intervals from each 100~ps simulation.}\label{fig:Li3N}
\end{figure*}

For the 9 non-argyrodite-type materials (see Figs.~\ref{fig:melt}f--j), the Vanilla method degrades the performance of SevenNet-0 due to catastrophic forgetting. In particular, the Vanilla-tuned model exhibits a significant loss of accuracy for certain materials. For instance, unlike DFT, the MLIP model yields spurious nitrogen diffusion in Li$_3$N (see Fig.~\ref{fig:Li3N}a), as well as short Li–Li bonds, as shown in the radial distribution function (RDF) plot in Fig.~\ref{fig:Li3N}b. We notice that the Replay-tuned MLIP also gives rise to unphysical Li–Li bonds during MD simulations. When examining DFT energies along the MD trajectories generated by each fine-tuned model (see Fig.~\ref{fig:Li3N}c), we find that, in addition to the Vanilla-tuned model, the Replay-tuned model energetically favors configurations with short Li--Li bonds more than DFT does.
In contrast, the MLIP fine-tuned with the reEWC method demonstrates remarkable consistency with DFT results for Li$_3$N. The contrasting results between the Replay and reEWC methods indicate that the unphysical Li–Li bond formation observed in the former is attributable to the relatively large parameter shift during fine-tuning (see Fig.~\ref{fig:loss}b). This underlines the importance of applying appropriate constraints on parameter updates to preserve generalizability and ensure stable MD simulations.

\section{Discussion}\label{sec3}

In Table~1, we compare the four fine-tuning methods based on the results discussed above. Although the Vanilla method improves accuracy for a specific target system, it suffers from severe catastrophic forgetting, resulting in poor performance on materials falling outside the training domain. Therefore, it is not a suitable choice for continual learning, where preserving the generalizability of the original MLIP is crucial. On the other hand, Replay mitigates the forgetting issue by periodically revisiting samples from the original training set. As a result, it significantly improves the performance of the pretrained MLIP on the target system and, to some extent, on other materials through knowledge transfer. However, its effectiveness depends on the composition of the Replay set. Moreover, because Replay does not explicitly constrain parameter updates, it carries the risk of degrading MD performance for materials with chemistries that differ substantially from the target system, as demonstrated in the case of Li$_3$N. While Replay generally incurs higher computational cost and memory usage compared to Vanilla, our mini-batch strategy demonstrates that the method remains feasible for lab-scale computational resources (e.g., 1 NVIDIA RTX A5000 GPU).  

EWC constrains changes to important parameters identified through the FIM, effectively preserving the generalizability of the pretrained model. It improves the accuracy of the pretrained MLIP for both the target system and other materials, with a low risk of failure in MD simulations. However, since EWC does not directly leverage data from the original training set during fine-tuning, it would result in slightly higher errors, particularly for non-target systems, compared to Replay and reEWC. 

reEWC, the hybrid fine-tuning method proposed in this study, achieves significant synergistic effects by combining Replay and EWC, effectively overcoming the limitations of each individual approach. It demonstrates excellent performance across all key aspects—including forgetting prevention, learning of target systems, and preservation of generalizability. Notably, the training cost of reEWC is comparable to that of Replay, indicating its high practical feasibility. Therefore, reEWC represents a promising fine-tuning strategy well-suited for continual learning of pretrained MLIPs.

Lastly, we comment on the selection of fine-tuning sets. As discussed throughout this work, forgetting-aware fine-tuning methods such as reEWC facilitate effective knowledge transfer, improving predictive accuracy even for materials outside the fine-tuning domain. This is a key advantage, enabling the development of universal MLIPs with minimal additional effort. To maximize this benefit, it is essential to sample configurations where the pretrained MLIP exhibits low predictive accuracy. In this study, we constructed the fine-tuning set using AIMD trajectories of LPSC at 600 and 1000 K, allowing us to capture diverse and complex PES features associated with Li migration in argyrodite-type SSEs. In contrast, if the fine-tuning set were generated solely from 600 K AIMD simulations of Li$_6$PS$_5$I, a material with significantly lower Li diffusivity~\cite{LPSI}, the resulting fine-tuned MLIPs suffer from pronounced PES stiffening, leading to underestimated Li diffusivities (see Fig. S8). On the other hand, although this study focuses on a single material for generating the fine-tuning dataset, multiple materials can be selected as fine-tuning targets, provided that the computational cost does not increase significantly. Such an augmentation of fine-tuning targets can further broaden the coverage of the fine-tuned MLIPs.

In conclusion, we proposed reEWC, a hybrid fine-tuning strategy that integrates Replay and EWC to effectively address catastrophic forgetting in pretrained MLIPs during the fine-tuning process. By fine-tuning the SevenNet-0 model on LPSC, we demonstrated that reEWC significantly improves accuracy on the target system, resolving PES softening and reproducing Li diffusivity in close agreement with DFT. Moreover, reEWC preserves the generalizability of the original model and enables knowledge transfer to a wide range of chemically diverse systems, including other argyrodite and non-argyrodite Li SSEs. While Replay and EWC individually contribute to forgetting prevention, each has notable drawbacks—Replay may impair MD stability for out-of-domain materials, and EWC alone is less effective in reducing energy and force errors. In contrast, reEWC achieves a favorable balance between stability and plasticity, combining their complementary strengths with minimal additional computational cost. Overall, reEWC offers a robust and effective solution for the continual learning of pretrained MLIPs, revolutionizing materials research by enabling large-scale and high-throughput simulations with accurate and transferable models.

\clearpage
\begin{table}
\caption{Overall comparison of fine-tuning methods}
\centering
\begin{tabular}{l|l|cccc}
\toprule
\textbf{Evaluation Criterion} & \textbf{Evaluation Basis} & \textbf{Vanilla} & \textbf{Replay} & \textbf{EWC} & \textbf{reEWC} \\
\midrule
Forgetting prevention & Loss on sMPtrj & Poor & \makecell{Moderate\\(depends on the Replay set)} & Good & Excellent \\
\midrule
\makecell[l]{Accuracy on\\target system} & Loss on LPSC FT set & Excellent & Excellent & Excellent & Excellent \\
\midrule
Generalizability & MD stability & Poor & \makecell{Moderate\\(unstable outliers)} & Good & Excellent \\
\midrule
Training cost & CPU elapsed time & $\times$1 & $\times$2 & $\times$1 & $\times$2 \\
\bottomrule
\end{tabular}
\end{table}
\clearpage

\section{Methods}\label{sec4}

\subsection{Theoretical details about EWC}

EWC originates from a Bayesian inference framework, where the objective is to find model parameters that perform well on both the pretraining dataset \(\mathcal{D}_{\text{pre}}\) and the fine-tuning dataset \(\mathcal{D}_{\text{FT}}\)~\cite{Regularization_EWC}. This means that our task is to maximize the posterior distribution \(p(\theta | \mathcal{D}_{\text{pre}}, \mathcal{D}_{\text{FT}})\) over the model parameters after observing both datasets, which reflects the confidence in the parameter values given all available evidence from both training phases.

The posterior distribution can be expressed using Bayes' rule as:
\begin{align}
p(\theta | \mathcal{D}_{\text{pre}}, \mathcal{D}_{\text{FT}}) = \frac{p(\mathcal{D}_{\text{FT}} | \theta) \cdot p(\theta | \mathcal{D}_{\text{pre}})}{p(\mathcal{D}_{\text{FT}} | \mathcal{D}_{\text{pre}})}.
\end{align}
Here, \(p(\mathcal{D}_{\text{FT}} | \theta)\) represents the likelihood of the fine-tuning data given the current parameters, while \(p(\theta | \mathcal{D}_{\text{pre}})\) serves as the posterior from the pretraining phase, which acts as a prior for the fine-tuning stage. Ignoring the constant denominator, maximizing the posterior is equivalent to minimizing the following negative log-posterior objective:
\begin{align}
\mathcal{L}(\theta) = -\log p(\mathcal{D}_{\text{FT}} | \theta) - \log p(\theta | \mathcal{D}_{\text{pre}}).
\label{log_loss}
\end{align}
The first term \(-\log p(\mathcal{D}_{\text{FT}} | \theta)\) corresponds to the fine-tuning loss, as it measures how well the current parameters \(\theta\) explain the fine-tuning data—lower likelihood indicates higher prediction error. The second term \(-\log p(\theta | \mathcal{D}_{\text{pre}})\) serves as a regularizer that encourages the model to retain knowledge from pretraining by penalizing parameter values that are unlikely under the pretrained posterior distribution.

Direct computation of the pretraining posterior, \(p(\theta | \mathcal{D}_{\text{pre}})\), is intractable for deep networks. Therefore, it is first expanded by using a second-order Taylor series around the pretrained parameters \(\theta_{\text{pre}}\):
\begin{align}
\log p(\theta | \mathcal{D}_{\text{pre}}) 
&\approx \log p(\theta_{\text{pre}} | \mathcal{D}_{\text{pre}}) \nonumber \\
&\quad + \frac{1}{2} (\theta - \theta_{\text{pre}})^\top H (\theta - \theta_{\text{pre}}).
\label{taylor}
\end{align}
Here, \(H\) denotes the Hessian of the negative log-prior evaluated at \(\theta_{\text{pre}}\), which describes the local curvature of the loss landscape around the pretrained solution.

Due to the computational complexity of calculating the full Hessian, it is approximated as the Fisher information matrix within the EWC framework. Under standard regularity conditions, the FIM can be expressed as the expected outer product of gradients:
\begin{align}
F_{ij} = \mathbb{E}_{\mathcal{D}_{\text{pre}}} \left[ \left( \frac{\partial}{\partial \theta_i} \log p(\mathcal{D}_{\text{pre}} | \theta) \right) \left( \frac{\partial}{\partial \theta_j} \log p(\mathcal{D}_{\text{pre}} | \theta) \right) \right],
\end{align}
where \( F_{ij} \) is the \( (i,j) \) component of the FIM. This formulation provides a practical and intuitive measure of the posterior’s curvature, indicating the sensitivity of the inference to parameter variations.

Practically, EWC adopts a diagonal approximation of the FIM by computing only the \(F_{ii}\) elements, thereby reducing memory requirements and computational overhead. The expectation in the FIM definition is also replaced with an empirical average over the pretraining dataset. Furthermore, since both the negative log-likelihood and the loss function quantify the model’s fitting quality on a given dataset, the log-likelihood term is substituted with the loss function. As a result, the FIM is expressed as shown in Equation~\ref{FIM}. On the other hand, the EWC loss in Equation~\ref{ewc_loss} is derived by: (1) converting the first term in Equation~\ref{log_loss} into the loss function evaluated on the fine-tuning dataset, (2) substituting Equation~\ref{taylor}—where the Hessian is approximated by the FIM—into Equation~\ref{log_loss}, and (3) introducing a hyperparameter \(\lambda\) to control the relative weight of the regularization term. 

\subsection{Fine-tuning of MLIP}

The fine-tuning dataset includes AIMD trajectories of \ch{Li6PS5Cl} at 600 and 1000 K for the 50\% Cl@4c configuration, where Cl@4c refers to the percentage of Cl atoms randomly occupying the 4c sites, with the remaining sites occupied by S atoms (see Fig.~S2). AIMD simulations use the NVT ensemble with a Nosé–Hoover thermostat \cite{nose}, a time step of 2 fs, and a total simulation time of 100 ps. The structures are sampled at 200 fs intervals. The dataset is randomly divided into training and validation subsets at a 9:1 ratio.

Fine-tuning the pretrained MLIP starts from the SevenNet-0 parameters and train for 400 epochs using the Adam optimizer, regardless of the fine-tuning methods. The learning rate is controlled using a cosine annealing scheduler with linear warm-up to prevent abrupt parameter shifts during fine-tuning; the learning rate increases gradually from 0 to 0.0001 over the first 50 epochs and subsequently decays to 0 over the remaining 150 epochs. Two cycles are repeated for 400 epochs of the fine-tuning process.

We employ the Huber loss function, consistent with the training of SevenNet-0. This loss function penalizes small and large residuals differently: it behaves quadratically for small residuals and linearly for large residuals beyond a threshold $\delta$, as defined below:

\begin{eqnarray}
\mathcal{L}_{\rm{Huber}}(y, \hat{y}, \delta) = \left\{ 
\begin{array}{ll}  
\frac{1}{2}(y - \hat{y})^2, & |y - \hat{y}| \le \delta, \\  
\delta \cdot \left(|y - \hat{y}| - \frac{1}{2}\delta\right), & \mathrm{otherwise}  
\end{array} 
\right.
\label{eq_Huber}
\end{eqnarray}

The $\delta$ is set to 0.01. The total loss \(\mathcal{L}\) combines normalized contributions from energy, force, and stress as follows:

\begin{eqnarray}
\mathcal{L} & = & \frac{1}{M} \sum_{i=1}^{M} \mathcal{L}_{\mathrm{Huber}} 
\left(\frac{E_i}{N_i}, \frac{\hat{E}_i}{N_i}, \delta \right) \nonumber \\
& & + \frac{\lambda_{F}}{3M \sum_{i=1}^{M} N_i} 
\sum_{i=1}^{M} \sum_{j=1}^{N_i} \sum_{k=1}^{3} 
\mathcal{L}_{\mathrm{Huber}} \left(F_{i,j,k}, \hat{F}_{i,j,k}, \delta \right) \nonumber \\
& & + \frac{\lambda_{S}}{6M} \sum_{i=1}^{M} \sum_{l=1}^{6} 
\mathcal{L}_{\mathrm{Huber}} \left(S_{i,l}, \hat{S}_{i,l}, \delta \right).
\label{eq_total_loss}
\end{eqnarray}
Here, \(M\) is the batch size and \(N_i\) is the number of atoms in sample \(i\). The values \(E_i\), \(F_{i,j,k}\), and \(S_{i,l}\) denote the DFT energy, the \(k\)-th Cartesian component of the force on atom \(j\), and the \(l\)-th component of the stress tensor, respectively. Their predicted values from a MLIP model are \(\hat{E}_i\), \(\hat{F}_{i,j,k}\), and \(\hat{S}_{i,l}\). All quantities are in units of eV for energy, eV/\AA\ for force, and kbar for stress. The coefficients \(\lambda_{F} = 1\) and \(\lambda_{S} = 0.01\) control the relative weight of force and stress losses.

We determine the regularization strength $\lambda$ for EWC and reEWC by balancing performance between the fine-tuning target and the pretrained domain. To this end, we first explore several values and ultimately set $\lambda = 10^6$ for EWC and $\lambda = 10^5$ for reEWC. These values yield sufficiently low losses on both the fine-tuning validation and sMPtrj datasets. For instance, EWC and reEWC achieve total losses below 0.1 on the validation set for the fine-tuning target, corresponding to RMSEs of approximately 1 meV/atom for energy, 70 meV/\AA\ for force, and 1.5 kbar for stress. On the much larger sMPtrj dataset (more than 1,600 times larger than the validation set), the corresponding total losses are only 0.87 and 0.58, respectively. Variations in the $\lambda$ values lead to only marginal changes in the results, rarely affecting our conclusions. The relationship between $\lambda$ values and loss metrics is summarized in the error bar plots in Fig.~S9.

When calculating the FIM, we focus on a subset of the pretrained dataset that exhibits relatively high accuracy, allowing for more effective identification of parameters critical to preserving pretrained performance. This subset comprises 1,154,095 configurations with Huber losses below a predefined threshold of 0.000726. 

\subsection{Calculation details for MD simulations}
In this study, all DFT calculations are performed using the Vienna Ab initio Simulation Package (\texttt{VASP}) \cite{VASP} with projector-augmented-wave (PAW) pseudopotentials~\cite{PAW}. The Perdew–Burke–Ernzerhof (PBE) \cite{PBE} functional is employed to describe the exchange–correlation energy. The cutoff energy and {\bf k}-point grid are determined to ensure convergence criteria are met, specifically an energy difference within 10 meV/atom, atomic forces within 0.02 eV/\AA, and stress components within 10 kbar. We perform structural relaxation for all materials until the atomic forces converge below 0.04 eV/\AA. For materials known to have cubic symmetry and intrinsic structural disorder in the atomic arrangement, such as argyrodite compounds and LLZO, the cubic cell shape is retained throughout the relaxation. All MLIP-based relaxations are conducted using the Atomic Simulation Environment (\texttt{ASE}) package \cite{ase}.

In AIMD simulations, we adopt $\Gamma$-point sampling and use the relaxed atomic structures from prior static calculations. NVT simulations are conducted with a Nosé–Hoover thermostat, a time step of 2 fs, and a total simulation duration of 100 ps. For systems containing hydrogen, such as \ch{LiBH4}, a shorter time step of 1 fs is used to accurately capture the dynamics of lighter atoms. For MLIP-based MD simulations, we use the \texttt{LAMMPS} \cite{lammps} package. The cell size is chosen such that the cell dimensions are approximately 10 $\rm \AA$ to minimize self-interaction under periodic boundary conditions.

\subsection{Structure generation for test dataset}

For argyrodite-type materials in the test dataset, we consider compounds with the general formula \ch{Li_{24+$x$}$M$4S20$X$4}, where \( M \) denotes a cation species, such as Si, Ge, Sn, P, As, or Sb, or their combinations, and \( X \) represents a halide anion (Cl, Br, or I) or their mixture. The value of \( x \) is determined to satisfy charge neutrality, depending on the specific cationic composition. To generate initial structures, 24 Li atoms are placed at the 24g Wyckoff positions, while, if available, the excess Li atoms are randomly distributed over the 48h and 16e atomic sites~\cite{arg_doped}.

Among non-argyrodite-type materials, several compositions, such as LLZO, LATP, LGPS, LYC, and LYB, are known to exhibit partial site occupancies. For each composition, we first generate fifty distinct configurations and identify the ten lowest-energy structures based on energies computed using the pretrained SevenNet-0 model. Subsequently, we select the final structure by comparing the DFT energies of these ten candidates and choosing the one with the lowest energy.

\subsection{Diffusivity calculation}
We calculate the self-diffusion coefficient of Li ions at 800, 1000, and 1200 K based on the MSD obtained from NVT MD simulations for 100 ps. The MSD is computed using the time-averaged method \cite{statistical}. Specifically, for a given time window $\tau$, the MSD is calculated as:
\begin{equation}
    \mathrm{MSD}(\tau) = \left\langle \frac{1}{N} \sum_{i=1}^{N} \left| \mathbf{r}_i(t+\tau) - \mathbf{r}_i(t) \right|^2 \right\rangle
\end{equation}
where $N$ is the number of Li ions, and $\mathbf{r}_i(t)$ represents the position vector of $i$-th Li ion at time $t$. The self-diffusion coefficient $D$ is then determined by fitting the data using the Einstein relation~\cite{einstein}:
\begin{equation}
    D = \frac{1}{6t} \mathrm{MSD}(t)
\end{equation}
Following previous literature~\cite{statistical}, we exclude the initial portion of the data where the MSD is below 4.5$\rm \AA^2$, as this corresponds to the ballistic regime dominated by vibrational motion of ions. Additionally, we omit data beyond 70\% of the total simulation time, where statistical reliability decreases due to limited ensemble averaging. To further improve statistical robustness, we perform five independent MD simulations with randomly assigned initial velocities for each system and report the averaged diffusivity.

\subsection{Weighted mean absolute error calculation}

In Fig.~\ref{fig:melt}, we construct a two-dimensional histogram to illustrate the consistency between DFT and MLIP quasi-melting ratios. To quantify the degree of agreement, we compute a WMAE, which emphasizes frequently observed error patterns. Specifically, let \(x_i\) and \(y_i\) denote the DFT and MLIP quasi-melting ratios, respectively, for the \(i\)-th composition--temperature pair, and let \(N_i\) represent the number of occurrences observed for that specific combination. The weight assigned to each point is then calculated as
\begin{equation}
w_i = \ln(N_i + 1),
\end{equation}
and the resulting WMAE is given by
\begin{equation}
\mathrm{WMAE} = \frac{\sum_i w_i \cdot |x_i - y_i|}{\sum_i w_i}.
\end{equation}
The WMAE provides a physically interpretable measure of model error, where values close to zero indicate near-perfect alignment between MLIP and DFT-predicted melting trends.

\section{Data availability}
The datasets generated and analyzed for this study, including the fine-tuning dataset, MD simulation trajectories, and result files, are available on Zenodo at https://doi.org/10.5281/zenodo.15686940.

\section{Code availability}
The code for SevenNet-reEWC is available at https://github.com/kskjs1203/SevenNet-reEWC.

\section{Acknowledgements}\label{sec5}
This research was supported by the Nano \& Material Technology Development Program through the National Research Foundation of Korea (NRF) funded by Ministry of Science and ICT (RS-2024-00407995), and the KIST Institutional Program (Project No. 2E33861).
The computations were carried out at Korea Institute of Science and Technology Information (KISTI) National Supercomputing Center (KSC-2025-CRE-0110) and at the Center for Advanced Computations (CAC) at Korea Institute for Advanced Study (KIAS).

\section{Author contributions}
J.K., J.L., S.K., and Y.K. conceived the initial idea. J.K. implemented the software for the overall SevenNet-reEWC framework, with S.O. making major contributions to the Replay methodology. Y.P. refactored and reviewed the entire implementation. J.K. and J.L. developed the MLIPs, performed simulations. J.K., J.L., S.K., and Y.K. mainly prepared the manuscript. S.H.$^1$ and S.H.$^2$ offered significant insight and guidance throughout the project. S.K and Y.K supervised the project in its entirety, encompassing conceptual development, experimental design, and data analysis. All authors contributed to discussions and approved the paper. (S.H.$^1$: Seungwoo Hwang, S.H.$^2$: Seungwu Han)

\bibliography{sn-bibliography}

\end{document}